\documentclass[aps,prl,showpacs,showkeys,twocolumn,amsmath,amssymb]{revtex4}

\usepackage{graphicx}
\usepackage{dcolumn}
\usepackage{natbib}
\usepackage{color,calc}
\usepackage{bm}
 
\begin{document}

\title{Pressure-induced enhancement of superconductivity and 
superconducting-superconducting transition in CaC$_6$}
\author{A. Gauzzi$^{1,4}$}
\email{andrea.gauzzi@upmc.fr}
\author{S. Takashima$^1$}
\author{N. Takeshita$^2$}
\author{C. Terakura$^2$}
\author{H. Takagi$^{1,2}$}
\author{N. Emery$^3$}
\author{C. H\'{e}rold$^3$}
\author{P. Lagrange$^3$}
\author{G. Loupias$^4$}

\affiliation{$^1$Department of Advanced Materials Science, University of Tokyo, 
Kashiwa, Chiba 277-8581, Japan\\
$^2$Correlated Electron Research Center (CERC), National Institute of Advanced 
Industrial Science and Technology (AIST), Tsukuba, Ibaraki 305-8562, Japan\\
$^3$Laboratoire de Chimie du Solide Min\'{e}ral-UMR 7555, Universit\'{e} Henri 
Poincar\'{e} Nancy I, B.P. 239, 54506 Vandoeuvre-l\`{e}s-Nancy Cedex, France\\
$^4$Institut de Min\'{e}ralogie et de Physique des Milieux Condens\'{e}s-UMR 7590, Universit\'{e} Pierre et Marie Curie, 4, place Jussieu, 75252, Paris, France}

\date{\today}

\begin{abstract}
We measured the electrical resistivity, $\varrho(T)$, of superconducting CaC$_6$ at ambient and high pressure up to 16 GPa. For $P \leq$8 GPa, we found a large increase of $T_c$ with pressure from 11.5 up to 15.1 K. At 8 GPa, $T_c$ drops and levels off at 5 K above 10 GPa. Correspondingly, the residual $\varrho$ increases by $\approx$ 200 times and the $\varrho(T)$ behavior becomes flat. The recovery of the pristine behavior after depressurization is suggestive of a phase transition at 8 GPa between two superconducting phases with good and bad metallic properties, the latter with a lower $T_c$ and more static disorder.

\end{abstract}


\maketitle
The interest in Graphite Intercalated Compounds (GICs) has been recently renewed after the discovery of superconductivity with much higher $T_c$ values than previously reported. The highest values hitherto reported are 6.5 and 11.5 K in YbC$_6$ and CaC$_6$, respectively \cite{wel,eme1}. Magnetic penetration depth measurements \cite{lam} and \emph{ab initio} calculations \cite{maz,cal} point at a conventional BCS $s$-wave scenario with a moderate electron-phonon coupling, $\lambda\approx 0.83$, and 2$\Delta/T_c \approx 3.6$, where $\Delta$ is the superconducting gap. Contrary to a simple-minded picture of doped graphene layers, the above calculations indicate that the intercalant plays a non trivial role in the pairing. For example, the relevant electronic states in CaC$_6$ have a marked hybrid Ca-C character with a 40\% Ca component \cite{cal}.

These new results raise the questions of the maximum $T_c$ that can be reached in GICs and of the role of the intercalant in enhancing $T_c$. In order to address these issues, in this letter we report on electrical resistivity, $\varrho(T)$, measurements in bulk superconducting CaC$_6$ at ambient and high pressure up to 16 GPa. To our knowledge this is the first study of the transport properties on CaC$_6$, for previously the superconducting transition was investigated only by means of magnetization measurements \cite{wel,eme1}.

Our results show that, for $P \leq$ 8 GPa, pressure increases $T_c$ in CaC$_6$ up to 15.1 K, the highest value hitherto observed for GICs. At 8 GPa, we find evidence of a structural phase transition that leads to a strong $T_c$ reduction and to bad metal properties. Our data analysis indicates that the $T_c$ increase is caused by an enhancement of $\lambda$. The latter is found to increase anomalously near 8 GPa, which accounts for the observed structural instability. Thus, even higher $T_c$ values might be obtained in CaC$_6$, provided the instability can be prevented.

For this work, we measured three bulk CaC$_6$ samples of size $\approx$1 mm $\times$ 0.4 mm $\times$ 0.2 mm prepared from platelets of $c$-axis oriented pyrolithic graphite, as described elsewhere \cite{eme2}. The $\varrho(T)$ measurements were carried out in a conventional four-probe bar configuration using a dc method. Due to the high reactivity of CaC$_6$, the electrical contacts were made using silver paste in a glove box. The samples were subsequently protected by a layer of halogen-free cryogenic grease to enable handling in air. Owing to their platelet shape and $c$-axis orientation, the in-plane $\varrho$ was measured.   

The ambient pressure measurements prior and after pressurization were carried out in a commercial PPMS system. The high-pressure study was carried out using a cube-anvil press enabling the four-probe measurement of electrical resistivity under nearly hydrostatic conditions up to 16 GPa down to 2.5 K, as described in detail elsewhere \cite{mor}. The sample is mounted within teflon capsule filled with a fluorinert liquid used as pressure-transmitting medium. Each run of measurements was carried out at constant pressure on both cooling and heating by adjusting the load.

We first present and analyze the ambient pressure results. The $\varrho(T)$ curve is shown in Fig. 1. One notes the low values of room temperature and residual resistivities, respectively $\varrho_{300K}$=46 $\mu\Omega$cm and $\varrho_0$=0.8 $\mu\Omega$cm, the quite large residual resistivity ratio, $RRR \equiv \varrho_{300K} / \varrho_0$=58, and the pronounced linear $T$ dependence in a wide temperature region above 150 K. For our superconducting samples, $\varrho_0$ is defined as the $\varrho$ value at the onset of the transition. Taking into account the $\sim 10\%$ uncertainty in the geometry of the contacts, the $\varrho_{300K}$ value reported here agrees well with the value reported for single crystal whiskers of pure graphite \cite{bac}. Above $T_c$, the $\varrho(T)$ curve was fitted by the generalized Bloch-Gruneisen formula \cite{gri}, where the electron-phonon coupling function $\alpha^2 F(\omega)$ was approximated by the following discrete decomposition into Einstein modes of energy $\omega_k$ \cite{lor}:
\begin{equation}
\alpha^2F(\omega)=\sum_{k}\alpha_k^2F_k \delta(\omega-\omega_k)
\label{eq:modes}
\end{equation}
The data of Fig. 1 are fitted using two modes of energies $\omega_1$=136 K and $\omega_2$=600 K and relative weights $\alpha_k^2F_1$=0.29 and $\alpha_k^2F_2$=0.71. The quality of the fit is so good that the fitting curve is indistinguishable from the experimental curve. Both energies and weights are in excellent agreement with the calculations of Calandra and Mauri \cite{cal}. Neither Eq. (1) with a single Einstein mode nor a restricted Bloch-Gruneisen formula with $\alpha^2 F(\omega) \sim \omega^4$ account for the data. Thus, no definite power-law behavior, e.g. $\varrho \sim T^5$, is seen at low temperatures. The low $\varrho_0$ value and the absence of saturation of $\varrho(T)$ at high temperatures suggest a small amount of static disorder and a large mean free path, well beyond the Ioffe-Regel limit.

The inset of Fig. 1 shows in detail the low-temperature region of the $\varrho(T)$ curve. One notes a sharp superconducting transition with onset and zero-resistance values of 11.5 and 11.0 K respectively. The former value coincides with the onset value of previous magnetization data \cite{eme1}, so we conclude that the sample is homogeneous and the resistive transition is not of percolative type. The sharpness of the transition further supports the conclusion of sample homogeneity. 
\begin{figure}
\includegraphics[width=7.8 cm]{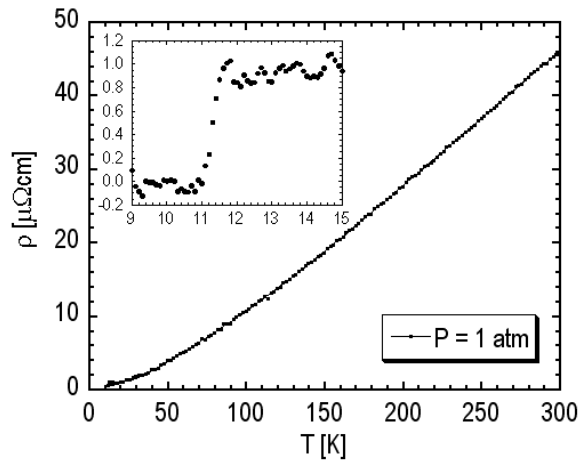}
\caption{In-plane resistivity of CaC$_6$ at ambient pressure. Note the linear behavior above 150 K, the low residual resistivity and the sharp superconducting transition at 11.5 K. The Bloch-Gruneisen fit described in the text is indistinguishable from the experimental curve.}
\label{rho_ambient}
\end{figure}
\begin{figure}
\includegraphics[width=7.8 cm]{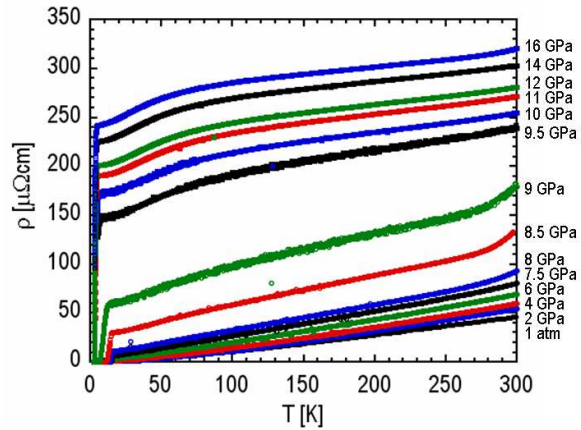}
\caption{(color online). Pressure-dependence of the in-plane resistivity of CaC$_6$. Note the sudden increase of residual resistivity above 8 GPa accompanied by a drop of $T_c$ down to 5 K (see Figure 3).}
\label{rho_curves}
\end{figure}
In summary, the excellent agreement between the experimental electron-phonon coupling function within the Bloch-Gruneisen model and the theoretical calculations of Ref. \cite{cal} in addition to the above magnetic penetration depth study \cite{lam} give clear evidence of conventional BCS superconductivity for CaC$_6$.

The results of the high pressure $\varrho$ measurements are shown in Figs. 2 and 3. Two distinct regimes below and above 8 GPa are found. In the former, the $\varrho(T)$ curves preserve the features of the ambient pressure curve of Fig. 1. In particular, the behavior remains that of a conventional metal with a markedly linear $T$-dependence at sufficiently high temperature and low $\varrho_0$. By analyzing the data, the following variations with pressure are observed (see Fig. 4): 1. $\varrho_0$ increases roughly quadratically with $\varrho_0$=0.8 and 8 $\mu \Omega$cm at 1 atm and 8 GPa respectively; 2. the room temperature resistivity, $\varrho_{300K}$, and the resistivity coefficient, $d\varrho/dT$, increase linearly, with $\varrho_{300K}$=46 and 94 $\mu \Omega$cm at 1 atm and 8 GPa respectively. 3. $T_c$ increases linearly with a large rate $\approx$0.5 K/GPa and reaches the maximum of 15.1 K at 7.5 GPa.

The correlation between $T_c$ and $d\varrho / dT$ suggests that the $T_c$ increase is caused by a pressure-induced enhancement of the electron-phonon coupling. To verify this possibility, we applied the previous spectral analysis of electron-phonon coupling function to the pressure-dependent $\varrho(T)$ curves. The result is summarized in Fig. 5. We indeed find a linear increase with pressure of the coupling constant for the 136 K mode, followed by an anomalous increase near 8 GPa. At 8.5 GPa, the coupling constant is 73\% larger than at 1 atm. This increase is accompanied by a mode softening from 136 K at 1 atm down to 85 K at 8.5 GPa. No significant variations of coupling constant are observed for the 700 K mode. However, this mode exhibits a hardening up to $\approx$1100 K near 8 GPa.
\begin{figure}
\includegraphics[width=7.8 cm]{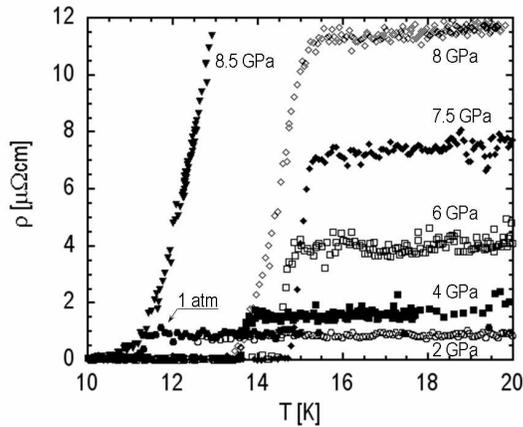}
\caption{Details of the superconducting transitions shown in Fig. 2. Note the transition broadening above 7.5 GPa.}
\label{zoom}
\end{figure}
\begin{figure}
\includegraphics[width=7.8 cm]{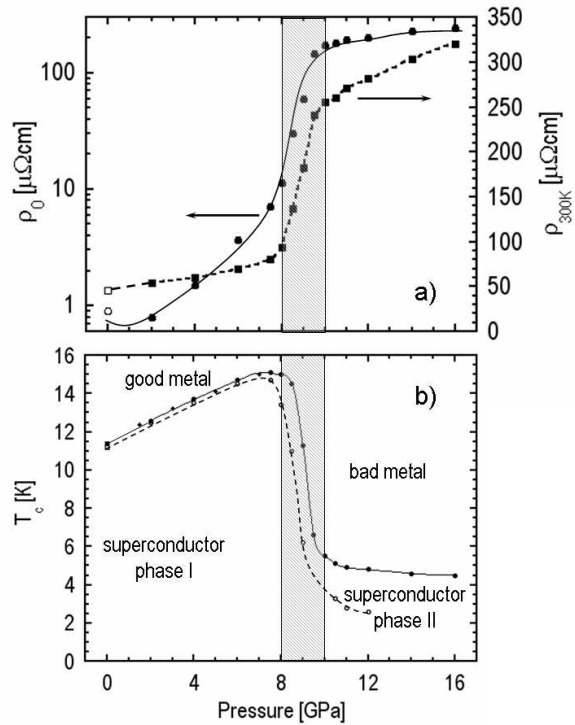}
\caption{Pressure dependence of $\varrho_0$ and $\varrho_{300K}$ (a) and of $T_c$ (b). The dashed area denotes the pressure range where large variations are observed. Lines are a guide to the eye. In (a), open symbols refer to sample no. 2 measured only at 1 atm. In (b), full and open symbols and full and broken lines refer to respectively onset and offset values. Different symbols refer to three different samples.}
\label{summary}
\end{figure}

The anomaly of both phonon modes near 8 GPa clearly indicates an incipient structural instability. In Fig. 3a one notes that $\varrho_0$ suddenly increases up to nearly 200 $\mu\Omega$cm, i.e. more than 200 times the ambient pressure value. This change corresponds to a pronounced flattening of the $\varrho(T)$ curves accompanied by an incipient downward curvature at low temperatures followed by an indication of localization below 10 K. Indeed, the above Bloch-Gruneisen model does no longer account for the behavior of these curves and the previous spectral analysis of the electron-phonon coupling function is no longer applicable. This drastic change of regime in the normal state transport is concomitant to a sudden drop of $T_c$ down to 5 K. The correlation between the $\varrho$ increase and the $T_c$ drop is apparent by comparing Figs. 4a and 4b. In the 10-16 GPa region, no further significant changes are observed. In particular, $\varrho_0$ continues to increase, although slowly, and the flattening of the $\varrho(T)$ curve becomes more pronounced. Interestingly, the $T_c$ onset values levels off at 5 K, although the offset values continue to slowly decrease. At 14 GPa, the zero-resistance state is not yet achieved at 2.5 K. In Fig. 4b, it is noted that the transition width, calculated as the difference between onset and offset values, remains narrow ($\leq$ 0.5 K) below 8 GPa. A large broadening is observed in the 8-10 GPa region, where $T_c$ drops. The maximum broadening, as large as 6 K, is found at 9 GPa, i.e. exactly at the midpoint of the $T_c$ drop. At larger pressure values, the transition sharpens back, although not completely, and the width is less than 2 K at 10.5 GPa.
\begin{figure}
\includegraphics[width=7.8 cm]{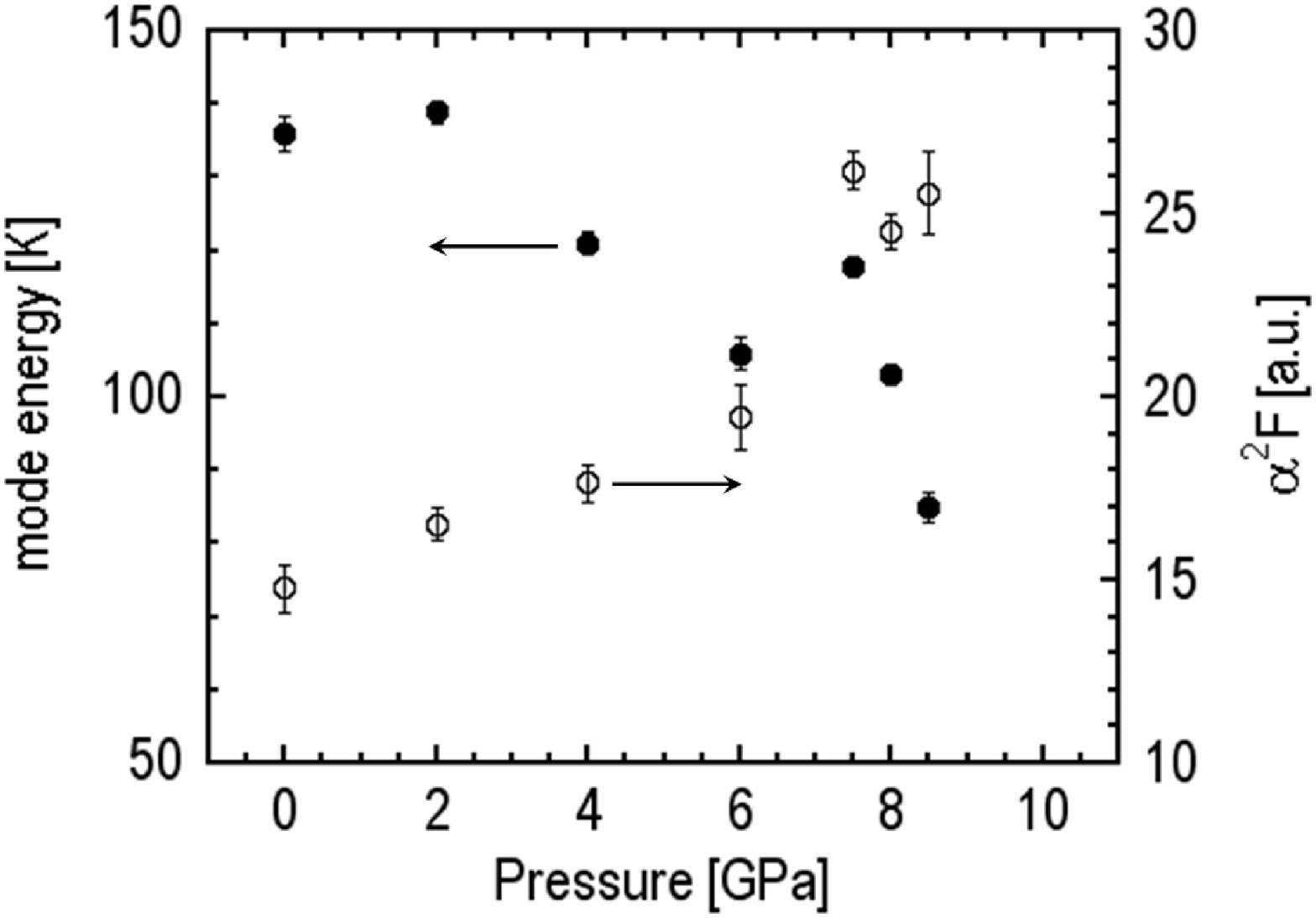}
\caption{Pressure-dependence of the energy and electron-phonon coupling constant for the 136 K phonon mode. Note the anomaly of the coupling near 8 GPa.}
\label{modes}
\end{figure}

The above results give evidence of a phase transition in the 8-10 GPa region. The transition broadening is an indication that fluctuations or a $T_c$ distribution are present. A further evidence of phase transition is provided by the almost full recovery of the pristine $\varrho(T)$ behavior after depressurization (data not shown). Thus, extrinsic effects like pressure-induced micro-cracks can not be invoked to account for the large changes of resistivity and of superconducting properties observed. After depressurization, both $\varrho_0$ and $\varrho_{300K}$ are higher than before pressurization. However, the difference is much smaller than the large change observed at 8-10 GPa. Specifically, we note the following. 1. After depressurization, $\varrho_0$ is larger by a factor of 8 only, to be compared with the factor of 200 of the 10 GPa data. 2. The overall behavior of $\varrho(T)$, is recovered after depressurization, although the $RRR$ value is smaller (17 instead of 58) because of the larger $\varrho_0$. 3. The superconducting transition of the pristine sample is fully recovered and both $T_c$ and transition width are unchanged.

An \emph{in situ} diffraction study under high pressure is required to unveil the phase transition at 8-10 GPa. At this point, we limit ourselves to note the following. 1. An incipient structural instability at $\approx$ 7 GPa, i.e. in good agreement with the experimental value, is predicted by \emph{ab initio} calculations of phonon dynamics \cite{cal2}. 2. One possible phase transition would be a modification of stacking sequence induced by pressure. Indeed, the different sequences found in the $M$C$_6$ family differ by a small amount of energy \cite{eme2}. 3. The large increase of $\varrho_0$ gives evidence of a large amount of static disorder in the high pressure phase. This disorder may arise from a disordered rearrangement of the intercalant atoms or by a spatially incoherent buckling of graphene layers. This would strongly reduce the mean free path of the charge carriers, thus accounting for the flattening of the $\varrho(T)$ curves observed experimentally. As a result, the transport regime would change from free electron-like into diffusive. Diffusive transport would be detrimental to superconductivity, in agreement with our experimental observation.

In summary, we reported the first $\varrho(T)$ measurements on bulk superconducting CaC$_6$ at ambient and high pressure up to 16 GPa. For $P \leq$ 8 GPa, we found that $T_c$ increases linearly with pressure up to 15.1 K with a large rate of 0.5 K/GPa. At 8-10 GPa, we gave evidence of a phase transition between a good and well ordered metal with relatively high $T_c$ and a bad disordered metal with lower $T_c$. Our quantitative analysis of the pressure-dependent $\varrho(T)$ data accounts for the $T_c$ increase in terms of a pressure-induced enhancement of the electron-phonon coupling. The anomaly of both relevant phonon modes near 8 GPa accounts for the structural instability leading to the bad metallic phase at high pressure. The present work suggests that still higher $T_c$ values are possible in GICs, provided the structural instability at 8 GPa can be avoided. To examine this possibility, further studies are needed to unveil the structural properties and the role of disorder on the normal state and superconducting properties of the high-pressure phase. 
\begin{acknowledgments}
This work is partly supported by CREST-JST and Grant in Aid for Scientific Research MEXT Japan. The authors thank M. Calandra, M. d'Astuto, F. Mauri, M. Nohara and A. Shukla for useful discussions. This work is dedicated in memory of J. Evetts.
\end{acknowledgments}

\end{document}